\documentclass[prb,superscriptaddress,aps,twocolumn,10pt,showkeys]{revtex4-1}
\usepackage{graphics,graphicx,epsfig,color,latexsym,float,amsmath,bm}
\usepackage[colorlinks,citecolor=blue,linkcolor=blue,urlcolor=blue]{hyperref}
\usepackage[version=4]{mhchem}
\usepackage{wrapfig}

\graphicspath{{figures/}}

\usepackage{ulem}
\usepackage{lipsum}
\usepackage{placeins}
	
\newcommand{\ket}[1]{\left| #1 \right>} % for Dirac bras
\begin{document}
\title{Spin-hedgehog-derived electromagnetic effects in itinerant magnets}

\author{G.~V. Paradezhenko}
\affiliation{Skolkovo Institute of Science and Technology, Moscow 121205, Russia}
\author{A.~A. Pervishko}
\affiliation{Skolkovo Institute of Science and Technology, Moscow 121205, Russia}
\author{N. Swain}
\affiliation{MajuLab, CNRS-UCA-SU-NUS-NTU International Joint Research Unit IRL 3654, Singapore}
\affiliation{Centre for Quantum Technologies, National University of Singapore 117543, Singapore}
\author{P. Sengupta}
\affiliation{School of Physical and Mathematical Sciences, Nanyang Technological University 637371, Singapore}
\author{D. Yudin}
\affiliation{Skolkovo Institute of Science and Technology, Moscow 121205, Russia}

\date{\today}

\begin{abstract}
In itinerant magnets, the indirect exchange coupling of Ruderman--Kittel--Kasuya--Yosida type is known to stabilize incommensurate spin spiral. Whereas an account of higher order spin interactions favors the formation of a noncoplanar magnetic texture. This is manifested by the finite Berry phase the conduction electrons accumulate when their spins follow this texture, leading thus to the topological Hall effect. We herein utilize the effective spin model with bilinear-biquadratic exchange interactions for studying the formation of the magnetic hedgehog lattice, that represents a periodic array of magnetic anti- and monopoles and has been recently observed in the B20-type compounds, in a three-dimensional itinerant magnet. As opposed to widely used Monte Carlo simulations, we employ a neural-network-based approach for exploring the ground state spin configuration in a noncentrosymmetric crystal structure. Further, we address the topological Hall conductivity, associated with nonzero scalar spin chirality, in the itinerant magnet due to the coupling to the spin hedgehog lattice, and provide the evidence of magneto-optic Kerr effect.
\end{abstract}

\maketitle

\section{Introduction} Recent achievements in information and communication technology stimulated the progress in various areas of artificial intelligence, putting forward even higher requirements for the hardware realization~\cite{Kim2020,Markovic2020,Christensen2022}. And magnetic systems are widely considered nowadays in the context of energy-efficient storage technologies~\cite{Fong2016,Verma2016,Back2020,Dieny2020,Shao2021}. Moving noncoplanar spin textures, {\it e.g.}, skyrmions, have been proposed lately as promising candidates to realize logic and memory devices with low current density~\cite{Jonietz2010,Schulz2012,Pereiro2014,Jiang2015,Woo2016,Koumpouras2016,Jiang2017,Buttner2017,Yu2017,Koumpouras2018,Yu2020,Pervishko2022,Wang2022}. The formation and stabilization of these textures is attributed to the competition between bilinear symmetric and antisymmetric magnetic exchange interactions~\cite{Bogdanov1989,Bogdanov2001,Rossler2006,Muhlbauer2009,Yu2010,Yu2011,Nagaosa2013,Yudin2017,Dohi2022} with the latter originating from relativistic effects, like spin-orbit coupling~\cite{Moriya1960}. 

The field equations of Dzyaloshinskii's original model~\cite{Dzyaloshinskii1958} allow soliton-type solutions that destroy collinear magnetic ordering~\cite{Bogdanov1989} with the tendency to the stabilization of spatially modulated textures with fixed sense of rotation. This is often manifested as a modulation of the spin arrangements with periodicity incommensurate to that of the lattice~\cite{Rossler2011}. Typically, the existence of these states as well as the mechanism of phase transformations by their nucleation is described in terms of continuous models whose magnetic energy allows Lifshitz invariants~\cite{Rossler2011}. In two-dimensional magnetic materials, the spin spiral background becomes unstable towards the formation of a skyrmion crystal upon applying an external magnetic field~\cite{Rossler2011,Leonov2015,Leonov2016}, whose stability is guaranteed by the topological reasoning~\cite{Bogdanov1995}. In three dimensions, an even richer variety of topological magnetic textures are stabilized~\cite{Zheng2018,Borisov2020,OHK20,Gobel2021,Kent2021}. A particular example is the lattice of magnetic hedgehogs that represents a noncoplanar magnetic structure with a periodic array of magnetic monopoles and antimonopoles~\cite{Kanazawa2016,OHK20,Okumura2022}. 

\begin{figure*}[ht!]
\centering
\includegraphics[width=1.0\textwidth]{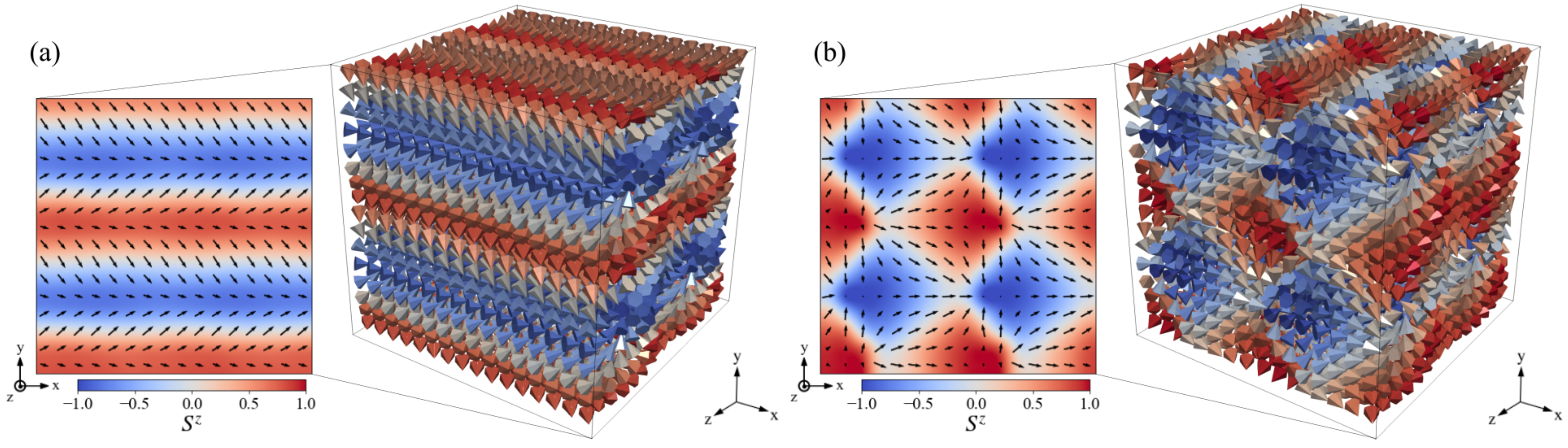}
\caption{\label{fig:config} Real-space spin textures as calculated by machine learning optimization of the model~\eqref{d-ham} in a cubic lattice $16\times16\times16$ in the absence of an external field. Note that the lattice spacing is set to unity throughout the calculations. A vertical spiral represents the ground state at $K/J=0.45$ and $D/J=0.3$ (a), whereas a hedgehog lattice is stabilized at $K/J=0.7$ and $D/J=0.3$ (b). Each cone depicts the spin orientation, whilst their $z$-components are highlighted by the color. Two-dimensional slices of the spin textures are taken normal to $\hat{\mathbf{e}}_z$, where the in-plane $(S_i^x,S_i^y)$ components of localized spins are depicted by black arrows in the $xy$ plane, and the out-of-plane $S_i^z$ components are visualized with the colorbar.}
\end{figure*}

Studying of skyrmions in two-dimensional chiral metallic magnets have unveiled their strong impact on electron transport properties~\cite{Lee2009,Neubauer2009,Kanazawa2011,Tome2021,Hayashi2021,SSP21}, and more generally has opened up an avenue towards the magnetic topological materials~\cite{Bernevig2022,Wang2022a}. Particularly, the topological Hall effect, one of the most striking properties of these systems, was shown to occur in the response to nonuniform and noncoplanar magnetization. Nonzero scalar spin chirality endows it with the properties distinct from those of the anomalous Hall effect that stems from the magnetic interaction between the localized and itinerant electrons even for collinear ordering. In magnetic topological materials, topological Hall effect arises when the electrons move through a noncoplanar magnetic texture background with rather strong magneto-electric coupling. The electrons accumulate finite Berry phase as their spins are constrained to be aligned with the local magnetization. We herein present evidence of novel magneto-transport and magneto-optic phenomena in electron systems coupled to a magnetic hedgehog lattice in a three-dimensional itinerant magnet. We start with the following spin Hamiltonian that was shown to host the spin hedgehog lattice ground state~\cite{OHK20}:
\begin{eqnarray}
    H_d & = & 2\sum_{\eta} \biggl[ -J \mathbf{S}_{\mathbf{Q}_{\eta}} \cdot \mathbf{S}_{-\mathbf{Q}_{\eta}} 
    + \frac{K}{N} \left(\mathbf{S}_{\mathbf{Q}_{\eta}} \cdot \mathbf{S}_{-\mathbf{Q}_{\eta}} \right)^2  \nonumber \\
    & - & i \mathbf{D}_{\eta} \cdot \left( \mathbf{S}_{\mathbf{Q}_{\eta}} \times \mathbf{S}_{-\mathbf{Q}_{\eta}} \right) \biggr] 
    - \sum_i \mathbf{h} \cdot \mathbf{S}_i, 
    \label{d-ham}
\end{eqnarray}
where $\mathbf{S}_{\mathbf{Q}} = \frac{1}{\sqrt{N}} \sum_{i} \mathbf{S}_i e^{-i\mathbf{Q} \cdot\mathbf{r}_i}$ is the Fourier transform of a localized spin $\mathbf{S}_i$, residing at site $i$ of a simple cubic lattice with the total number of $N$ sites. In the following, $\mathbf{S}_i$'s form a classical vector field of unit length. Note that Eq.~\eqref{d-ham} arises as an effective spin Hamiltonian upon integrating out conduction electrons in an $sd$-exchange model of itinerant electrons in a cubic lattice, coupled to localized moments, with additional spin-orbit coupling~\cite{Akagi2012,Hayami2014,Hayami2017,OHK20}. 
The first term represents the Ruderman--Kittel--Kasuya--Yosida-reminiscent exchange coupling of the strength $J$ that originates from the second-order expansion with respect to $sd$-exchange interaction. Among a number of multi-spin interactions that result from the expansion, we keep only the next-dominant biquadratic exchange of the strength $K$ as defined by the second term. A subsequent expansion in powers of the spin-orbit coupling strength leads to the antisymmetric Dzyaloshinskii-Moriya interaction, specified by the vectors $\mathbf{D}_\eta$ (provided $|\mathbf{D}_\eta|=D$). The last term stands for the Zeeman coupling to an external magnetic field $\mathbf{h}$. 

In Eq.~\eqref{d-ham}, all the exchange interactions are long ranged in real space and defined by particular wave numbers $\mathbf{Q}_{\eta}$. This inherits the itinerant nature of electrons. Specifically, the wave vectors $\mathbf{Q}_{\eta}$ are set by the multiple maxima in the spin-dependent bare susceptibility of itinerant electrons~\cite{HOM17}. We restrict our analysis to a minimal set of orthogonal cubic wave vectors $\mathbf{Q}_1 = (Q,0,0)$, $\mathbf{Q}_2 = (0,Q,0)$ and $\mathbf{Q}_3 = (0,0,Q)$, that corresponds to the $3Q$ hedgehog lattice state. Following Ref.~\cite{OHK20}, we choose $Q = \pi/4$ (period of eight lattice sites) and assume $\mathbf{D}_{\eta} \parallel \mathbf{Q}_{\eta}$ that stabilizes proper-screw-type spin texture. The magnetic field $\mathbf{h}$ is applied along the [001], [110], and [111] directions, in line with Ref.~\cite{OHK20}. 

The rest of the Paper is organized as follows. In Sect.~\ref{sect:methodology}, we briefly describe a machine learning method utilized for searching ground states of the model Hamiltonian~\eqref{d-ham}. In Sect.~\ref{sect:hedgehog}, we explore the magnetic hedgehog lattice configuration in terms of its spin structure factor and scalar spin chirality. In Sect.~\ref{sect:cond}, the optical conductivity of a three-dimensional itinerant magnet where the conduction electrons are coupled to the spin hedgehog lattice via the magnetic exchange interaction is calculated based on linear response theory. In Sect.~\ref{sect:Kerr}, we address the magneto-optic properties of the itinerant magnet by quantifying the optical rotation. Sect.~\ref{sect:conclusion} summarizes our findings and provide a brief discussion on their possible experimental implications.

\section{Methodology}\label{sect:methodology} 
We use the recently developed neural network approach~\cite{Kwon2019,SSP21} to simulate the Hamiltonian~\eqref{d-ham} in a cubic lattice with $16\times16\times16$ sites under periodic boundary conditions. The method consists of training a fully connected neural network to search for ground state spin configurations. At each iteration, we generate a batch $\mathbf{X}$ of size $n_{\mathrm{b}}$ of $n$-dimensional normal random vectors and feed it to the neural network. The input features are then decoded as $\mathbf{Y} = \mathbf{XW} + \mathbf{C}$, where $\mathbf{W}$ is the $n \times 3N$ matrix of weights, $\mathbf{C}$ is the $1 \times 3N$ bias, and $N = L^3$ is the number of lattice sites. The $n_b \times 3N$ matrix $\mathbf{Y}$ of output features is then reshaped to form a batch of size $n_{\mathrm{b}}$ of three-dimensional spins configurations $(\mathbf{S}_1,\ldots,\mathbf{S}_N)$ on a lattice, while the normalization of output spins $\mathbf{S}_i$ to unit vectors serves as an activation function. 

\begin{figure*}[ht!]
\centering
\includegraphics[width=0.95\textwidth]{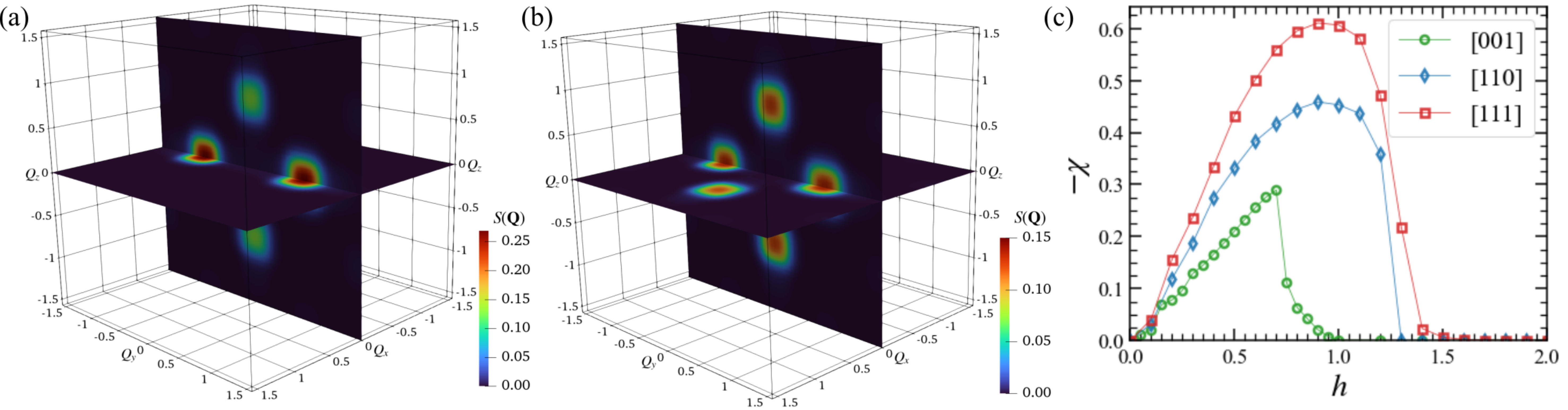}
\caption{\label{fig:struct}Static spin structure factor $S(\mathbf{Q})$ calculated by means of machine learning optimization for the vertical spiral (a) and spin hedgehog lattice (b). The magnitude of $S(\mathbf{Q})$ is depicted by color and visualized on the surface of two intersecting planes defined by $(Q_x,Q_y)$ and $(Q_y,Q_z)$. (c) Net spin chirality $\chi$ evaluated for the magnetic hedgehog lattice upon varying the magnitude of the external magnetic field applied in the [001], [110], and [111] directions.}
\end{figure*}

The neural network weights are trained to minimize the batch-averaged Hamiltonian $\langle H_d \rangle_{n_\mathrm{b}}$ that represents a cost function. We extract the Fourier transforms $\mathbf{S}_{\mathbf{Q}_{\eta}}$ associated with the set of wave vectors $\mathbf{Q}_{\eta}$, estimate the conjugates $\mathbf{S}_{-\mathbf{Q}_{\eta}} = \bar{\mathbf{S}}_{\mathbf{Q}_{\eta}}$, and substitute the result to Eq.~\eqref{d-ham}. In our simulations, the hyperparameters are $n_{\mathrm{b}} = 1024$ and $n=64$, we perform $10^4$ steps of the Adam optimizer~\cite{Adam2014} for training the weights with the learning rate $10^{-3}$. Upon completion of the training process, the neural network maps input features into the ground state spin configuration. We implement the method using GPU-based TensorFlow library for machine learning~\cite{TF}. 

\section{Magnetic hedgehog lattice}\label{sect:hedgehog}
Real-space magnetic states as obtained by machine learning optimization in the absence of the external field are shown in Fig.~\ref{fig:config}. The ground state evolves from a vertical spin spiral order (Fig.~\ref{fig:config}a) to a spin hedgehog lattice (Fig.~\ref{fig:config}b) with increasing $K$ at fixed $D$. A detailed study of the magnetic phase diagram of the model~\eqref{d-ham} with respect to $K$ and $D$ can be found in Ref.~\cite{OHK20}, we here focus on the magnetic hedgehog lattice exclusively. To identify different magnetic orderings and address the stabilization of a noncoplanar magnetic state, we calculate the static spin structure factor:
\begin{equation}\label{struct}
    S(\mathbf{Q}) = \frac{1}{N^2} \sum_{ij} \langle \mathbf{S}_i \cdot \mathbf{S}_j \rangle_{n_\mathrm{b}} e^{-i\mathbf{Q}\cdot (\mathbf{r}_i - \mathbf{r}_j)},
\end{equation}
where the average is taken over the batch of size $n_\mathrm{b}$. The results are shown in Fig.~\ref{fig:struct}. Expectedly, $S(\mathbf{Q})$ exhibits peaks at the wave vectors $\mathbf{Q}_\eta$ chosen in the model Hamiltonian~\eqref{d-ham}. In Fig.~\ref{fig:struct}a, the vertical spiral state represents a $2Q$ ordering, characterized by two pairs of symmetry-related peaks in $S(\mathbf{Q})$ at $\mathbf{Q} = \pm \mathbf{Q}_2$ (strong) and $\mathbf{Q} = \pm \mathbf{Q}_3$ (weak). As opposed, in Fig.~\ref{fig:struct}b, shown is the magnetic hedgehog state with a distinct six-peak structure in $S(\mathbf{Q})$ -- three pairs of symmetry related peaks at $\mathbf{Q} = \pm \mathbf{Q}_1$, $\pm \mathbf{Q}_2$, and $\pm \mathbf{Q}_3$ -- underscoring its $3Q$ nature. 

Noncoplanar arrangement of the spins in the hedgehog lattice makes this configuration a three-dimensional companion of a two-dimensional skyrmion lattice. Spin chirality at site $i$ is defined as a sum of triple products of the spins residing on triangular plaquettes enclosing $\mathbf{S}_i$:
\begin{equation}\label{chi-site-def}
    \chi^{\gamma}_{i} = \frac12 \sum_{\alpha,\beta,\nu_{\alpha},\nu_{\beta}} \epsilon^{\alpha\beta\gamma}
    \nu_{\alpha}\nu_{\beta} \, 
    \mathbf{S}_i \cdot \left( \mathbf{S}_{i + \nu_{\alpha}\mathbf{\hat{e}}_{\alpha}} \times \mathbf{S}_{i + \nu_{\beta}\mathbf{\hat{e}}_{\beta}} \right),
\end{equation}
where $\epsilon^{\alpha\beta\gamma}$ is the Levi-Civita symbol, $\alpha,\beta,\gamma=x,y,z$, and  
$\nu_{\alpha,\beta} =~\pm1$. Summing over all the sites and $\gamma$ and averaging over the batch of size $n_\mathrm{b}$, one can obtain the net spin chirality $\chi = N^{-1}\sum_{\gamma,i} \langle \chi_i^{\gamma} \rangle_{n_\mathrm{b}}$. Results for the net chirality are shown in Fig.~\ref{fig:struct}c, where we explore the evolution and eventual suppression of noncoplanar ordering upon applying an external magnetic field. Working with the $3Q$ hedgehog lattice parameters, we show the calculated chirality $\chi$ versus the magnitude $h$ of the external field along the [001], [110], and [111] directions in Fig.~\ref{fig:struct}c. The magnetic hedgehog lattice persists up to $h_c^{[001]}\simeq1.1$, $h_c^{[110]}\simeq1.3$ and $h_c^{[111]}\simeq1.5$ (in units of $t_1$) for the given field directions, respectively. Our findings based on machine learning optimization are in good agreement with the standard Monte Carlo simulations in Ref.~\cite{OHK20}, where the ground state phases were examined in detail.

\section{Optical conductivity}\label{sect:cond}
We are now in position to explore transport properties of itinerant electrons coupled to this noncoplanar spin texture via the following tight-binding Hamiltonian~\cite{OHK20}:
\begin{equation}
    H_\mathrm{tb}=-\sum\limits_{\langle i, j \rangle}c_i^\dagger(t_{ij}+i\lambda\mathbf{\sigma}\cdot\mathbf{d}_{ij})c_j^{}-J_K\sum\limits_i c_i^\dagger(\mathbf{S}_i\cdot\mathbf{\sigma})c_i^{}, \label{H-def}
\end{equation}
where $c_i=(c_{i\uparrow},c_{i\downarrow})$ denotes the electron annihilation operator at site $i$, the unit vector $\mathbf{d}_{ij}$ stands for neighboring sites of site $i$, and $\mathbf{\sigma} = (\sigma^x, \sigma^y, \sigma^z)$ are the Pauli matrices. The first term in Eq.~\eqref{H-def} contains the kinetic energy (with hopping strength $t_{ij}$) and spin-orbit interaction with coupling strength $\lambda$. Following Ref.~\cite{Hayami2017}, we consider hopping between neighboring ($t_1$) and third-neighboring ($t_3$) sites. The spin hedgehog lattice is coupled to the conduction electrons via $sd$-exchange coupling of the strength $J_K$, the last term. 

To address the optical conductivity, we apply the Kubo formula 
(see, {\it e.g.}, Refs.~\cite{Nagaosa2006,AM76}),
\begin{equation}\label{conductivity2}
    \sigma_{\alpha\beta}(\omega)=\frac{i\hbar}{L^3}\sum\limits_{m\neq n}
    \frac{f_n-f_m}{E_m-E_n}\frac{\langle n\vert j_\alpha\vert m\rangle\langle m\vert j_\beta\vert n\rangle}{E_n-E_m-\hbar\omega+i\eta},
\end{equation}
where $f_m$ represents the Fermi-Dirac distribution estimated at the energy $E_m$, specified by the single-particle state $\ket{m}=\vert i\sigma\rangle$ at $i$th site with spin $\sigma$, while $j_{\alpha}$ ($\alpha = x,y,z$) is the current density of the conduction electrons,

\begin{equation}\label{current}
    j_{\alpha} = -\frac{i e}{\hbar} \sum\limits_{\langle i,j \rangle_{\alpha}} \left[ c_i^{\dagger} 
    \left( t_{ij} + i \lambda \sigma^{\alpha} \right)
    c_{j}^{} - \mathrm{H.c.} \right],
\end{equation}
where the summation $\langle i,j \rangle_{\alpha}$ is carried out 
over the first and third nearest neighbors in the $\alpha$ direction.
The small broadening $\eta$ introduced to Eq.~\eqref{conductivity2} is associated with conduction electrons scattering off the localized magnetic moments.
In practice, we form a $2N\times 2N$ 
matrix of the single-particle Hamiltonian \eqref{H-def}
and diagonalize it using the LAPACK library for linear algebra 
to find its exact eigenvalues $E_m$ and eigenvectors $|m\rangle$~\cite{Shahzad2020,SSP21}.

\begin{figure*}[t]
\centering
\includegraphics[width=0.99\linewidth]{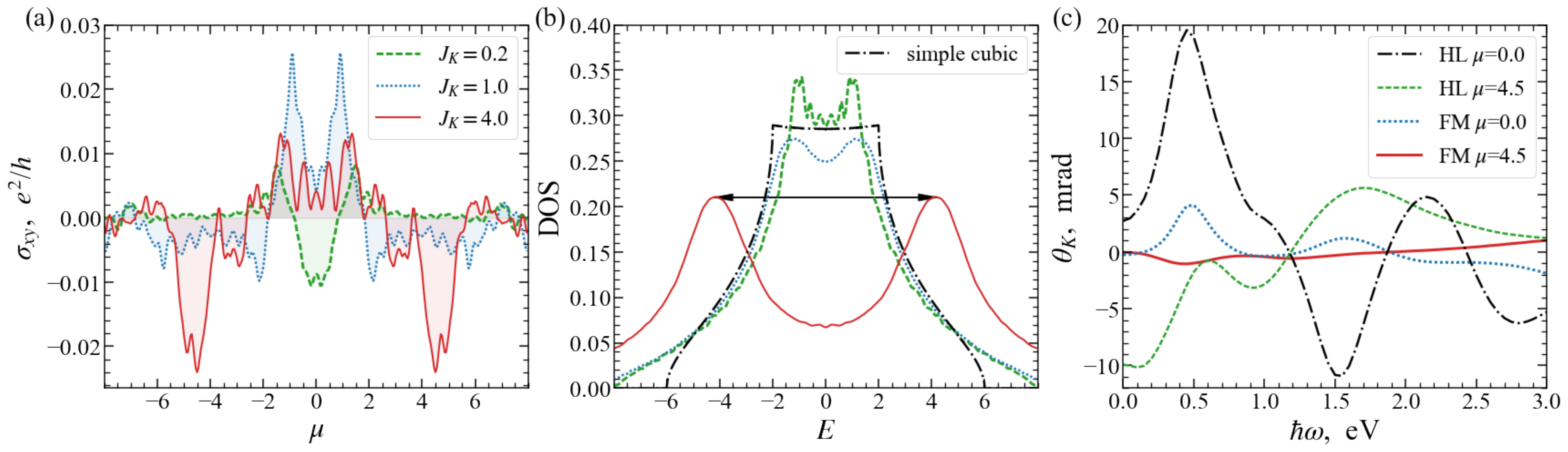}
\caption{\label{fig:cond} (a) Topological Hall conductivity $\sigma_{xy}$, evaluated from Eq.~\eqref{conductivity2} at $\omega=0$, as a function of the chemical potential $\mu$. (b) Density of states of the conduction electrons~\eqref{H-def} coupled to the magnetic hedgehog lattice (see Fig.~\ref{fig:config}b) in a $16\times16\times16$ cubic lattice, where exactly eight hedgehogs are present. The dash-dotted line visualizes the analytical formula~\eqref{dos-simple-cubic}. Note that a larger coupling $J_K$ gives a more pronounced subband splitting (marked by the black double sided arrow). Here, $\mu$, $J_K$, and $E$ are measured relative to $t_1$. (c) Frequency-dependent magneto-optic Kerr angle $\theta_K$, calculated using Eq.~\eqref{Kerr-angle} in the moderate coupling regime $J_K=4t_1$. The $sd$-exchange coupling to a magnetic hedgehog lattice (HL), as depicted in Fig.~\ref{fig:config}b, and to a collinear ferromagnet ordering (FM), as defined by $\mathbf{S}_i=\hat{\mathbf{e}}_z$ in~\eqref{H-def}, are shown.
Note that all the results are obtained at $T=0$.}
\end{figure*}

The Hall conductivity as evaluated according to Eq.~\eqref{conductivity2} at $\omega=0$ is shown in Fig.~\ref{fig:cond}a. In our numerical simulations, we assume $t_3/t_1 = 0.2$ and $\lambda/t_1 = 0.5$, and the broadening is set to $\eta/t_1 = 0.2$. Note that we restrict our analysis to $T=0$ and vary the $sd$-coupling strength $0.2t_1\leq J_K\leq4t_1$. Clearly, the spin degeneracy of a simple cubic crystal is lifted up by coupling the itinerant and localized subsystems. In fact, for a simple cubic lattice, the tight-binding model~\eqref{H-def} allows an exact analytical solution for the density of states~\cite{Jelitto1969,Modrak1979,Oppeneer92}:
\begin{equation}\label{dos-simple-cubic}
    \nu(E)=\frac{1}{\pi^3|t_1|
    }\int\limits_{u_1}^{u_2}
    \frac{d u}{\sqrt{1-u^2}} K\left[1 - \left(\frac{E}{4t_1} + \frac{u}{2}\right)^2 \right],
\end{equation}
on the condition that $\lambda=t_3=J_K=0$, while $u_1 = \max\left\lbrace-1,-2-E/(2t_1)\right\rbrace$ and $u_2 = \min\left\lbrace 1,2-E/(2t_1)\right\rbrace$. Here, $K(u)$ is the complete elliptic integral of the first kind. The density of states~\eqref{dos-simple-cubic}, shown by dash-dotted black line in Fig.~\ref{fig:cond}b, is an even function $\nu(-E)=\nu(E)$ being zero for $|E|>6t_1$. Switching on the coupling to the magnetic texture leads to the formation of two subbands separated by $2J_K$, as clearly visible in Fig.~\ref{fig:cond}b. The spin-orbit interaction has almost no influence on the topological Hall conductivity $\sigma_{xy}(\omega=0)$ as presented in Fig.~\ref{fig:cond}a, and the transverse charge current is generated by the magnetic exchange coupling to the spin hedgehog lattice.

\section{Magneto-optic Kerr effect}\label{sect:Kerr}
We further explore how the coupling to the magnetic hedgehog lattice affects the optical rotation as can be observed in a typical realization of magneto-optic Kerr effect. In practice, once a linearly polarized electromagnetic wave is reflected off the magnetic film, it becomes elliptically polarized with the rotation of the polarization principal axis being quantified by the Kerr angle,~$\theta_K$. Currently available experimental setups allow to capture the optical rotation with nanoradian accuracy~\cite{Gong2017}. Assume a linearly polarized electromagnetic wave $\mathbf{E}_\mathrm{inc}(z,t)\propto \hat{\mathbf{e}}_x e^{-i\omega t-i\omega z/c}$ of the frequency $\omega$ shines the itinerant magnet~\eqref{H-def} whose optical conductivity is specified by the expression~\eqref{conductivity2}. We restrict our analysis to normal incidence and consider the wave propagating along the $\hat{\mathbf{e}}_z$. The incident light is then partly reflected by the magnet, $\mathbf{E}_\mathrm{refl}\propto\{r_+(\hat{\mathbf{e}}_x+i\hat{\mathbf{e}}_y)+r_-(\hat{\mathbf{e}}_x-i\hat{\mathbf{e}}_y)\}e^{-i\omega t+i\omega z/c}$, and partly transmitted into the magnet. Note that circularly polarized modes with the refraction indices $r_\pm$ do not couple to each other upon reflection. More formally, the Kerr angle $\theta_K$ and ellipticity $\epsilon_K$ are specified by
\begin{equation}\label{eq:kerr}
    \tan\left(\epsilon_K+\frac{\pi}{4}\right)e^{-2i\theta_K}=\frac{r_+}{r_-}.
\end{equation}
In a typical experiment, $\theta_K,\epsilon_K\ll 1$, validating thus the approximation $\theta_K+i\epsilon_K\approx i(r_+-r_-)/(2r_-)$.

Consider a thin film of the itinerant magnet occupying the space $-L<z<0$, as specified by Eq.~\eqref{H-def} where the conduction electrons are coupled to the magnetic hedgehog lattice. Following the standard Fresnel approach, the refraction indices $r_\pm$ can be derived by matching interface conditions for electromagnetic fileds at $z=-L$ and 0 (see, {\it e.g.,} Ref.~\cite{Catarina2020}):
\begin{equation}
    r_\pm=\frac{n_\mp(1-\sqrt{\varepsilon})+i(n_\mp^2-\sqrt{\varepsilon})\tan(\omega n_\mp L/c)}{n_\mp(1+\sqrt{\varepsilon})-i(n_\mp^2+\sqrt{\varepsilon})\tan(\omega n_\mp L/c)}.
\end{equation}
Note, in this formula, we assume the half-space $z<-L$ is occupied with a material of the relative permittivity $\varepsilon$. For sufficiently thin films ($L\ll c/\omega$) one can relate $n_\pm^2=\varepsilon_0\{1+i\sigma_\pm/(\omega L)\}$ to the components of the conductivity tensor $\sigma_\pm=\sigma_{xx}\pm i\sigma_{xy}$ \eqref{conductivity2} with $\varepsilon_0$ standing for vacuum permittivity. If this is the case, the Kerr rotation angle and Kerr ellipticity~\eqref{eq:kerr} read
\begin{equation}\label{Kerr-angle}
    \theta_K+i\epsilon_K=\frac{2\pi\alpha\sigma_{xy}/\sigma_0}{(\pi\alpha\sigma_{xx}/\sigma_0+1)^2+(\pi\alpha\sigma_{xy}/\sigma_0+i)^2},
\end{equation}
where $\alpha$ is the fine-structure constant, the conductance quantum $\sigma_0=e^2/(4\hbar)$, and $\varepsilon=1$. Our numerical findings suggest that the effect is the most pronounced in the moderate coupling regime, {\it e.g.}, $J_K = 4t_1$, as shown in Fig.~\ref{fig:cond}c. Meanwhile, a close inspection of Fig.~\ref{fig:cond}c, unambiguously reveals that the assumption made ($\theta_K,\epsilon_K\ll1$) is justified. To provide a better understanding of the impact of the magnetic hedgehog lattice on the optical rotation, we also consider the coupling to a trivial ferromagnet state. Note that under normal incidence only polar magneto-optic Kerr effect with magnetization perpendicular to the reflection surface and parallel to the plane of incidence, {\it i.e.,}  $\mathbf{S}_i=\hat{\mathbf{e}}_z$ in~\eqref{H-def}, can be observed. In case of the spin hedgehog lattice as shown in Fig.~\ref{fig:cond}c, a higher value of the Kerr angle $\theta_K$ (up to 20 mrad) as compared to collinear state (less than 4 mrad) is clearly noticeable. Thus, the magneto-optic Kerr effect offers a feasible platform to assess the degree of spin noncoplanarity.

\section{Conclusions}\label{sect:conclusion}
In our study, we addressed a simple cubic itinerant magnet from the perspective of noncoplanar magnetic texture stabilization, accompanied by the study of its impact on magneto-transport and magneto-optic properties. The itinerant nature of the conduction electrons allows one to consider the system in terms of the effective spin model with bilinear-biquadratic long-range exchange interactions, where the magnetic spiral becomes unstable towards the formation of the spin hedgehog lattice. Complementing earlier studies, where the magnetic ground state had been assessed with the use of standard Monte Carlo routine and simulated annealing, we developed a machine-learning-based approach to explore the magnetic energy landscape and associated spin textures. Our numerical findings suggest that the multi-spin interaction is indeed responsible for the stabilization of the magnetic hedgehog lattice, showing up nonzero scalar spin chirality. We further addressed the topological Hall effect attributed to the nonzero scalar spin chirality by estimating off-diagonal {\it dc} conductivity and the magneto-optic Kerr effect upon the electromagnetic light reflection off the itinerant magnet with the spin hedgehog texture. Recently, spin hedgehog lattices have been identified in MnSi$_x$Ge$_{1-x}$~\cite{Fujishiro2019}. Upon varying Si/Ge substitution it is possible to observe phase transition among two-dimensional skyrmion lattice and three-dimensional magnetic hedgehog lattices, which may allow experimental verification of the results reported in this study.

\acknowledgements
We acknowledge use of the computational resources at the Skoltech supercomputer ``Zhores''~\cite{Zhores} in our numerical simulations. The work of A.A.P. was supported by the Russian Science Foundation Project No. 22-72-00021. P.S. acknowledges support from the Ministry of Education (MOE), Singapore, in the form of AcRF Tier 2 grant no. MOE2019-T2-2-119. D.Y. acknowledges the support from the Russian Science Foundation Project No. 22-11-00074.

%\FloatBarrier

\bibliographystyle{apsrev4-2}
\bibliography{draft.bbl}

\end{document}